# a deep learning-based ODE solver for chemical kinetics

Tianhan Zhang[1], Yaoyu Zhang[2*], Weinan E[3], and Yiguang Ju[1]

[1]*Department of Mechanical and Aerospace Engineering, Princeton University, Princeton, NJ, 08544, USA*

[2]*School of Mathematical Sciences, Institute of Natural Sciences, MOE-LSC, and Qing Yuan Research Institute, Shanghai Jiao Tong University, Shanghai, 200240, P.R. China*

[3]*Department of Mathematics, Program in Applied and Computational Mathematics, Princeton University, Princeton, NJ, 08544, USA*

**Developing efficient and accurate algorithms for chemistry integration is a challenging task due to its strong stiffness and high dimensionality. The current work presents a deep learning-based numerical method called DeepCombustion0.0 to solve stiff ordinary differential equation systems. The homogeneous autoignition of DME/air mixture, including 54 species, is adopted as an example to illustrate the validity and accuracy of the algorithm. The training and testing datasets cover a wide range of temperature, pressure, and mixture conditions between 750 – 1200 K, 30 – 50 atm, and equivalence ratio = 0.7 – 1.5. Both the first-stage low-temperature ignition (LTI) and the second-stage high-temperature ignition (HTI) are considered. The methodology highlights the importance of the adaptive data sampling techniques, power transform preprocessing, and binary deep neural network (DNN) design. By using the adaptive random samplings and appropriate power transforms, smooth submanifolds in the state vector phase space are observed, on which two three-layer DNNs can be appropriately trained. The neural networks are end-to-end, which predict temporal gradients of the state vectors directly. The results show that temporal evolutions predicted by DNN agree well with traditional numerical methods in all state vector dimensions, including temperature, pressure, and species concentrations. Besides, the ignition delay time differences are within 1%. At the same time, the CPU time is reduced by more than 20 times and 200 times compared with the HMTS and VODE method, respectively. The current work demonstrates the enormous potential of applying the deep learning algorithm in chemical kinetics and combustion modeling.**

1. Introduction

In recent years, due to global warming and air pollutions, there is increasing attention focused on climate and environmentally benign fuels such as biofuels [1], new combustion concepts, and technologies including low-temperature combustion chemistry [2][3]. The complex fuel molecular structures and the underlying complex physical and chemical processes in biofuel combustion need numerical simulations to solve detailed chemical kinetics efficiently. From a practical point of view, detailed chemical kinetics are usually modeled using reaction rates equations, a typical multi-scale system of ODEs. For example, the characteristic timescales of reactions vary from several seconds to picoseconds [4]. Simultaneously, the concentrations of hundreds of species range from ~$10^{-1}$ to less than $10^{-20}$. The inherent multi-scale features of the chemical kinetics render numerical algorithms challenging to develop. In previous studies, there are three main types of methods. The first one is the model reduction, such as computational singular perturbation [5], directed relation graph method [6], and path flux analysis [7]. The second one is the adaptive integration method, such as G-scheme [8] and hybrid multi-timescale method [9]. The third one is parameterization or tabulation methods, such as high dimensional model representations [10], in situ adaptive tabulation [11], and orthonormal polynomials [12]. However, even equipped with state-of-the-art numerical methods, large scale simulations using detailed chemistry still cost tens of millions of CPU hours [13]. The expensive computation on detailed chemistry proposes an urgent need for more efficient algorithms.

*zhyy.sjtu@edu.cn

On the other hand, with the rapid accumulating simulation data and the surprising effectiveness of DNNs, deep learning algorithms have shown high potentials in many scientific computing areas [14]. Neural networks have been applied in the combustion area for more than twenty years. Christo et al. started applying artificial neural networks (ANN) to model chemistry source term in 2D turbulent jet flames in 1996 [15,16], later Blasco et al. combines ANN with self-organizing-map (SOM) to improve the accuracy of the temporal prediction in 0D simulations[17,18]. J-Y Chen et al. used the existing in situ adaptive tabulation (ISAT) data to train ANNs [19]. Afterward, there are many different types of methods to train ANNs based on priori datasets, including flamelet tables[20–22], PDF tables[23], LEM LUT [24] etc. Recently, deep learning is also implemented into large eddy simulations to provide model parameters [25], select models [26], and optimize mechanism parameters [27].

However, there are still some challenging problems using ANNs to model chemistry kinetics in combustion, especially in modeling phenomena and intermediate species related to low-temperature chemistry, such as two-stage ignitions [22], OH and $HO_2$ radicals in the cool flame zones [28], and so on.

The main goal of the current work is to demonstrate the capability of the DNN as an efficient ODE solver for chemical kinetics in the homogeneous ignition cases. In the following section, the sampling techniques and the training procedures are introduced, focusing on the active sampling, preprocessing, and neural network design. Then, the accuracy and efficiency of the DNNs are demonstrated. Finally, conclusions are drawn.

## 2. Methodology

Dimethyl ether (DME), one of the simplest fuels with well-examined low-temperature chemistry, is chosen as the test fuel for the current study. A comprehensive reduced HP-Mech DME mechanism with 54 species and 418 reactions, as well as ignition and cool flame validation, is adopted [29]. The initial conditions are chosen as T = 750 – 1200 K, P = 30 – 50 atm, and equivalence ratio $\Phi$ = 0.7 – 1.5. The labeled data are generated by the in-house code ASURF+, which uses HMTS or VODE [30] method to integrate the chemistry and has been extensively validated [31,32].

This work aims at developing an end-to-end neural network for chemistry integration. Consequently, the input state vector is defined as $[X_{i=1,56}]^T \equiv [T, P, Y_{i=1,54}]^T$, where $X_i$ represents the state vector, $Y_i$ is the $i^{th}$ species mass fraction. The corresponding output gradient vector is $[R_{i=1,56}] \equiv [\Delta T, \Delta P, \Delta Y_{i=1,54}]^T / \Delta t$, where $R_i$ represents the temporal gradients defined as $R_i = \frac{X_i(t+\Delta t) - X_i(t)}{\Delta t}$. In this study, $\Delta t = 10^{-7}$ s.

The data sampling consists of two parts. The first part is to choose initial conditions of {T, P, $\Phi$} randomly, which is evenly distributed in the three-dimensional initial condition space $[T, P, \Phi] \in \mathbb{R}^3, T \in [750, 1200], P \in [30, 50], \Phi \in [0.7, 1.5]$. In total, 1100 ignition cases are simulated by the in-house code resulting in more than 100 GB raw data. In order to make a concise and representative training dataset, the second part of sampling is to decide which state vectors at different time points to be included. This selection is adaptive based on the heat release rate. In the end, 2.3 million data points are included. 90% of them are used as training data, 10% as testing data.

Before moving to the training step, it turns out that the preprocessing of the sampled raw data is necessary. The main reason is due to the wide range of scales of species concentrations and the information of minor species tends to be largely ignored by the neurons given their small magnitudes. Based on the Arrhenius law, the reaction rates are proportional to $\sim Y_i \exp\left(-\frac{T_a}{T}\right) = \exp(\ln Y_i - \frac{T_a}{T})$, where $T_a$ is the activation temperature. Considering temperature is one of the elements in the state vectors, from a dimensional analysis point of view, it is natural to consider using a logarithm-type transformation of mass fractions as the neural network's inputs. Besides, to avoid singularities when mass fractions are zero and have more flexibilities to adjust data distributions, power transformation $y_i = \frac{y^\lambda - 1}{\lambda} (\lambda > 0)$ is chosen as the preprocessing method for the mass fractions.

The corresponding phase diagram is shown in Figure 1, where $\lambda = 0.1$. 9 out of 56 entries of the state vectors are chosen to show the data distribution. By using the adaptive sampling and power transform, the majority of data points are distributed on the smooth submanifolds. Due to the Frequency-Principle of neural networks [33], the continuous and smooth submanifolds are the necessary foundations for proper training and generalization performance of neural networks for chemistry kinetics. Figure 1 also shows the magnitude separation of temporal gradients between the LTI < 1000 K and the higher temperature part, which leads to worsened uncertainties of DNN prediction in the low-temperature regime. The underlying reason is that the DNNs always try to learn the main features of the train samples first. In the current case, the DNN learns large magnitude temporal gradients in the high-temperature zone first, then tries to generalize the model to the low-temperature zone, which weakens its performance since the HTI and LTI are inherently different physics processes. As a result, another DNN was trained only using the data points below 1400 K to deal with the LTI.

Figure 2 shows the illustration of the structure of the DNNs. Each DNN has three hidden layers, with 1600, 400, 400 neurons, respectively. For each neuron, the activation function is ReLU. The loss function uses mean absolute function, which out-performed the mean square function in the training process. The binary neural network structure is not necessary but proves to be very useful. During the simulation, the choice of neural networks is solely based on the data point temperature. The trained DNNs can be easily implemented into any code since it only involves matrix multiplication. Even though deep learning techniques have been applied in numerous areas successfully, but it turns out to be a highly non-trivial task to adopt it into chemical kinetics of combustion with practical success.

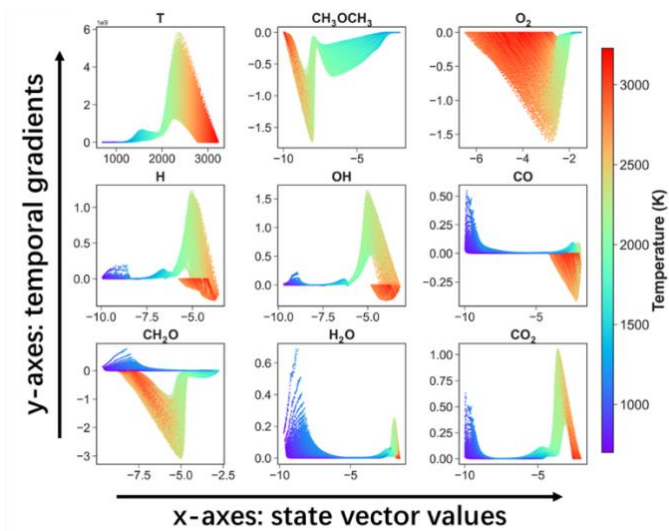

Figure 1 Phase diagrams of data points in the training dataset.

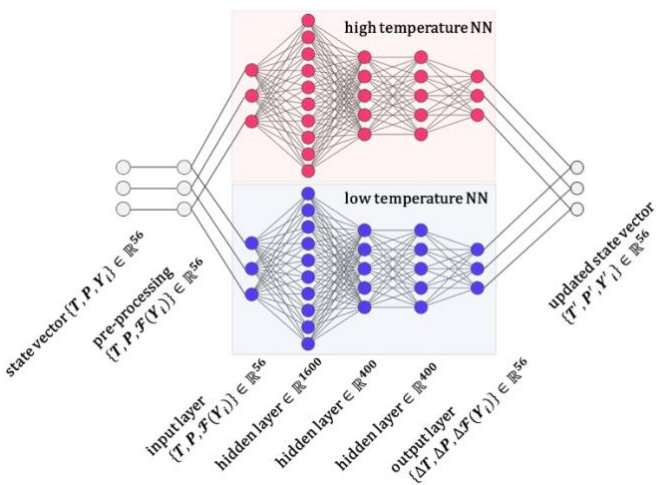

Figure 2 Illustration of the DNN structure

## 3. Results and discussion

Figure 3 shows the temporal evolution comparison between a previously developed efficient numerical method HMTS and the newly developed DNNs. The initial conditions are T = 800 K, P = 40 atm, equivalence ratio = 1.0, respectively. It shows that the prediction made by the DNN agrees very well with that of the HMTS method. Another 22 full results consisting of predictions in 56 dimensions with initial temperature T = 750 – 1200 K are included in the supplementary materials.

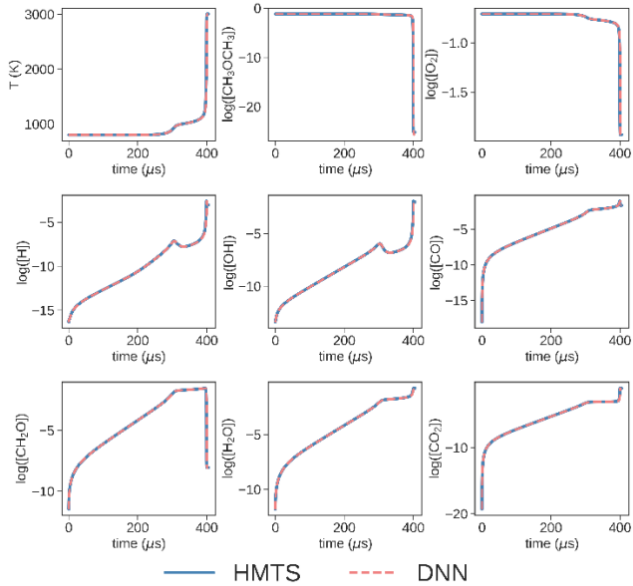

Figure 3 Temporal evolutions of temperature, pressure and species concentrations for the initial condition T = 800 K, P = 40 atm, equivalence ratio = 1.0.

Figure 4 shows the comparisons of the ignition delay time distribution and the CPU time, respectively. Here the ignition delay time is defined as the maximum heat release point or the temporal temperature rates. Figure 4a shows that for a wide range of temperature conditions, the HMTS and the DNN has a satisfying agreement in the ignition delay time calculation. The maximum difference is within 1%. At the same time, the CPU cost of DNN is significantly reduced compared with HMTS, which is already much faster than the VODE solver. For reference, both the VODE and the HMTS solvers are included to demonstrate the calculation efficiency. As seen in Figure 4b, HMTS is more than ten times faster than the VODE solver. However, the DNN is more than 20 times faster than HMTS, which is an extraordinary acceleration for chemistry integration.

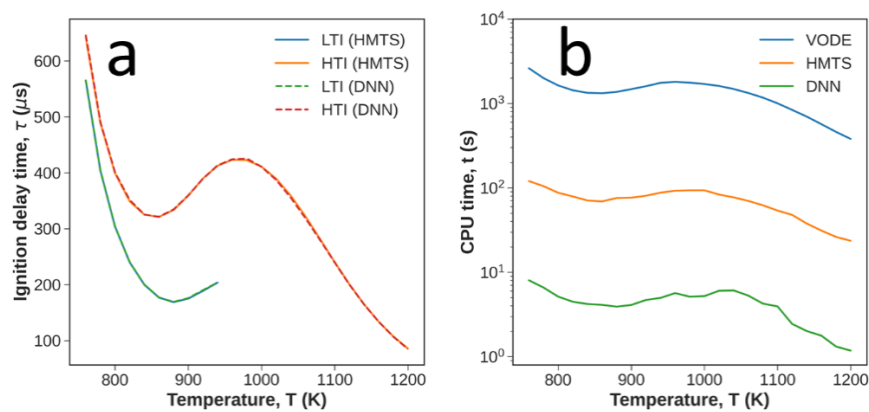

Figure 4 a: ignition delay time comparison; b: CPU time comparison

## 4. Conclusions

The present work developed and validated a new deep learning-based ODE solver for complex chemistry integration. As it is the very first step of a series of work applying deep learning algorithms into combustion simulations with detailed chemistry, the solver is named DeepCombustion0.0. The DNNs are trained using the labeled data generated by the in-house code ASURF+. A wide range of the initial conditions is covered by using the adaptive sampling based on the temporal rates of the state vector. After preprocessing species concentrations using power transform, the smooth submanifolds of training data points distributed in the phase diagrams are obtained. To accurately consider the drastic changes of the vector gradient in low-temperature chemistry and high-temperature chemistry, a binary DNN structure is used to build the solver. The results show that temporal evolutions of state vectors predicted by the DNN solver agree well with two previously used numerical methods, HMTS and VODE, and the predicted ignition delay time differences between these methods are within 1%. At the same time, the CPU time is reduced by more than 20 and 200 times compared to HMTS and VODE, respectively. The current work demonstrates the great potential of the deep learning algorithms in chemistry integration and combustion modeling.

## Acknowledgments

The authors would like to thank the grant support from the National Science Foundation with grant number DMS-1638352 and the Ky Fan and Yu-Fen Fan Membership Fund.